\documentclass[12pt]{amsart}
\usepackage{amsmath}
\usepackage{amssymb}
\usepackage[latin1]{inputenc}
\usepackage[T1]{fontenc}
\newtheorem{lemma}{Lemma}

\newtheorem{example}{Example}
\newtheorem{remark}{Remark}
\newtheorem{definition}{Definition}

\newtheorem{corollary}{Corollary}

\newcommand{\tr}{{\rm tr }}
\newcommand{\lh}{\mathcal{L(H)}}
\renewcommand{\th}{\mathcal{T(H)}}

\newcommand{\ket}{\rangle}

\newcommand{\C}{\mathbb{C}}

\newcommand{\N}{\mathbb{N}}

\newcommand{\R}{\mathbb{R}}

\newcommand{\be}{\begin{equation}}
\newcommand{\eeq}{\end{equation}}
\newcommand{\bet}{\begin{equation*}}
\newcommand{\eeqt}{\end{equation*}}
\newcommand{\bea}{\begin{eqnarray}}
\newcommand{\eeqa}{\end{eqnarray}}
\newcommand{\beat}{\begin{eqnarray*}}
\newcommand{\eeqat}{\end{eqnarray*}}

\newcommand{\h}[1]{\mathcal{#1}}
\newcommand{\hil}{\mathcal{H}}

\newcommand{\pa}{\partial}
\newcommand{\la}{\lambda}
\def\<{\langle}
\def\>{\rangle}

\begin{document}

\title{Density matrix reconstruction from displaced photon number distributions}

\author{Jukka Kiukas}
\address{Institute for Theoretical Physics, University of Hannover, Hannover, Germany}
\email{jukka.kiukas@itp.uni-hannover.de}
\author{Juha-Pekka Pellonp\"a\"a}
\address{Turku Centre for Quantum Physics, Department of Physics and Astronomy, University of Turku, Turku, Finland}
\email{juha-pekka.pellonpaa@utu.fi}
\author{Jussi Schultz}
\address{Turku Centre for Quantum Physics, Department of Physics and Astronomy, University of Turku, Turku, Finland}
\email{jussi.schultz@utu.fi}
\begin{abstract}
We consider state reconstruction from the measurement statistics of phase space observables generated by photon number states. The results are obtained by inverting certain infinite matrices. In particular, we obtain reconstruction formulas, each of which involves only a single phase space observable.

\end{abstract}
\maketitle

\section{Introduction}
A density operator of a quantum system is determined by any informationally complete set of measurements performed on the system; this means that the state is uniquely specified by the collective outcome statistics of such measurements (see, for instance \cite{Prugovecki, Busch}). However, this point of view is rather abstract; in practical applications one instead aims to derive explicit reconstruction formulas for the density operator in terms of the empirical distributions in question. Of course, informational completeness of the measurements is necessary for the existence of such reconstruction formulas. In quantum optics one typically uses the set of rotated quadratures \cite{Dariano1994, DAriano, Leonhardt1996}, which can easily be measured by homodyne detection \cite{Leonhardt}. Another option is to use phase space observables. In particular, one can reconstruct the density matrix from the collection of phase space observables generated by number states. The associated distributions are sometimes called displaced photon number distributions, and the entire collection is called \emph{photon number tomogram} by Manko \emph{et al.} \cite{Mancini1997, tomogram1, tomogram2}.

The purpose of this paper is to use the method of infinite matrix inversion, as in \cite{KiPeSc}, to derive state reconstruction formulas involving the measurement outcome distributions of phase space observables generated by the number states. We consider two different types of formulas, involving (1)  the entire tomogram, and (2) only a single observable. We find one formula of type (2) for each displaced photon number distribution; up to our knowledge, such formulas have previously been obtained only for the observable generated by the vacuum state.

The paper is organized as follows. In section \ref{preliminaries} we fix the notations and consider the informational completeness of phase space observables. The relevant results concerning the   inversion of infinite matrices are proved in section \ref{matrix}. Section \ref{suure} contains the main results of this paper. After presenting the basic properties of the phase space observables generated by the number states and discussing the possibility of measuring the observables, we prove two reconstruction formulas. In section \ref{discussion}, we discuss the advantages and disadvantages of the two reconstruction scenarios considered in this paper.

\section{Preliminaries}\label{preliminaries}
Let $\hil$ be a complex separable Hilbert space and
$\{|n\ket\mid n\in \N\}$ be an orthonormal basis of $\hil$
where $\N := \{0,1,2,\ldots\}$. 
The basis is identified
with the photon number basis, or Fock basis, in the case where $\hil$ is associated with the single mode
electromagnetic field. Let $a$ and $a^*$ denote the usual raising and lowering operators associated with the above
basis of $\hil$, and let $N=a^*a$ be the selfadjoint number operator. 
Now the phase shifting unitary operators are $R(\theta):=e^{i\theta N}$. Define the shift operator of the complex plane
$D(z)=e^{z a^*-\overline z a}$, $z\in\C$, for which the identities $D(z)^*=D(z)^{-1}=D(-z)$ and
$R(\theta)D(z)R(\theta)^*=D\big(ze^{i\theta}\big)$ hold. The matrix elements of $D(z)$ with respect to the number basis are
\begin{equation}\label{Dmatriisi}
\langle m\vert D(z)\vert n\rangle = (-1)^{\max\{0,n-m\}} e^{i\theta (m-n)}\sqrt{\frac{\min \{m,n\}!}{\max\{m,n\}!}} e^{-r^2/2} r^{\vert m-n\vert} L^{\vert m-n\vert}_{\min \{m,n\}} (r^2),
\end{equation}
where $z=re^{i\theta}$ and
$$
L^\alpha_s (x) :=\sum_{u=0}^s \frac{(-1)^u}{u!} \binom{s+\alpha}{s-u} x^u
$$
is the associated Laguerre polynomial.

Let $\lh $ be the set of bounded
operators on $\hil$, and $\th $ the set of trace class operators. We let $\|\cdot\|_1$
denote the trace norm of $\th $ and the operator norm of $\lh $ is denoted
by $\|\cdot\|$. When $\hil$ is associated with a
quantum system, such as the single mode electromagnetic field, the states of the system are represented by positive operators $\rho\in \th $ with the unit trace, density operators,
and each state is fully characterized by the matrix elements $\rho_{mn}:=\langle m\vert\rho \vert n\rangle$ with respect to the given basis. The observables are associated with the normalized positive operator measures (POMs) which, in the case of phase space observables, are defined on the Borel $\sigma$-algebra $\h B(\C)$ of subsets of $ \C\cong \R^2$.\footnote{A normalized positive operator measure, defined on a $\sigma$-algebra $\Sigma$ of subsets of a set $\Omega$, is a map $\mathsf{E}:\,\Sigma\to \lh $ which is $\sigma$-additive in the weak operator topology, and has the property $\mathsf{E}(\Omega)=I$ (the identity operator), 
that is, for which $X\mapsto\tr[\rho \mathsf{E}(X)]$ is a probability measure for each state $\rho$.} The measurement outcome statistics of a phase space observable $\mathsf{E}:\h B(\R^2) \to \lh $
in a state $\rho$ are given by the probability measure $X\mapsto \tr[\rho \mathsf{E}(X)]$.

For each positive operator $K$ of trace one, define the phase space POM $\mathsf{E}^K :\h B(\C)\to\lh $ by
\begin{equation}\label{kovarianttisuure}
\mathsf{E}^K(X):=\int_{X} D(z)KD(z)^* \frac{d^2 z}{\pi},
\end{equation}
where the integral exists in the $\sigma$-weak sense. This measure is covariant in the sense that
$$
D(\alpha) \mathsf{E}^K(X)D(\alpha)^*=\mathsf{E}^K (X+\alpha),
$$
for all $X\in\h B (\C)$ and $\alpha\in\C$. Furthermore, each covariant phase space observable is of the above form \cite{Holevo, Werner}. We use the notation $G^K$ for the operator density related to $\mathsf{E}^K$, that is, $G^K:\C\rightarrow \h L(\hil)$, $z\mapsto G^K( z) =D(z) KD(z)^*$.  
For a fixed density operator $\rho$, the phase
space probability measure associated
with $\mathsf{E}^K$, i.e. $X\mapsto\tr[\rho  \mathsf{E}^K(X)]$, has a density
$G^K_\rho:\C\to [0,\infty)$, given by
$$
G^K_\rho(z) = \tr[\rho D(z)KD(z)^*] =\tr[\rho G^K (z)].
$$
If $K$ is a one-dimensional projection, that is $K=\vert\psi\rangle\langle\psi\vert$ for some $\psi\in\hil$, $\Vert\psi\Vert =1$, we use the notations $\mathsf{E}^\psi:= \mathsf{E}^{\vert\psi\rangle\langle\psi\vert}$, $G^\psi:=G^{\vert\psi\rangle\langle\psi\vert}$ and $G^\psi_\rho:=G^{\vert\psi\rangle\langle\psi\vert}_\rho$ respectively.

When reconstructing the state of the system directly from some measurement data, the measured observables are required to distinguish between any two states:

\begin{definition}\rm A set $\h M$ of observables $\mathsf{E}:\h B(\Omega)\to \lh $ is
\emph{informationally complete}, if any two states $\rho$ and $\rho'$ are equal
whenever $\tr[\rho \mathsf{E}(X)] = \tr[\rho'\mathsf{E}(X)]$ for all $\mathsf{E}\in \h M$ and $X\in \h B(\Omega)$.
\end{definition}
In other words, the informational completeness of a set $\h M$ of observables means that the totality of
the corresponding measurement outcome distributions determines
the state $\rho$ of the system. Clearly, a set $\h M$ of observables is informationally complete
if and only if $\rho=0$ whenever $\rho$ is a selfadjoint trace class operator with $\tr[\rho \mathsf{E}(X)]=0$
for all $\mathsf{E}\in \h M$ and $X\in \h B(\Omega)$. If $\h M$ consists of a single observable $\mathsf{E}$, we say that $\mathsf{E}$ is an informationally complete observable. A covariant phase space observable $\mathsf{E}^K$ is known to be informationally complete if $\tr [KD(z)]\neq 0$ for almost all $z\in\C$ \cite{Prugo}. As a consequence of this, we get the following lemma, which shows that $\mathsf{E}^K$ is informationally complete whenever $K$ is a finite matrix. In particular, the observables generated by the number states are informationally complete. 
\begin{lemma}\label{infolemma}
Let $K$ be a positive operator with unit trace, whose matrix representation with respect to the number basis $\{ \vert n\rangle \mid n\in\N \}$ is finite. Then the covariant phase space observable $\mathsf{E}^K$ generated by $K$ is informationally complete.
\end{lemma}
\begin{proof}
Since the matrix representation of $K$ is finite, $K$ can be written as a finite sum $K=\sum_{m,n=0}^k K_{mn}\vert m\rangle\langle n\vert$. Due to the linearity of the trace, we then have 
$$
\tr [KD(z)] =\sum_{m,n=0}^k K_{mn}\langle n\vert D(z)\vert m\rangle
$$
for all $z\in\C$. According to equation \eqref{Dmatriisi}, we have $\langle n\vert D(z)\vert m\rangle=0$ exactly when 
$$
\vert z\vert^{\vert m-n\vert} L^{\vert m-n\vert}_{\min\{m,n\}} (\vert z\vert^2) =0,
$$
which is a polynomial of $\vert z\vert$ of order $m+n$. We thus find that  $\langle n\vert D(z)\vert m\rangle=0$ for only a finite number of points $z\in\C$, which then implies that
$$
\tr [KD(z)] \neq 0
$$
for almost all $z\in\C$. Hence, $\mathsf{E}^K$ is informationally complete.
\end{proof}

Consider now an arbitrary covariant phase space observable $\mathsf{E}^K$. Let $P_n$ be the projection onto the $n$-dimensional subspaces spanned by the vectors $\vert k\rangle$, $k=0,1,\ldots,n-1$, that is $P_n =\sum_{k=0}^{n-1} \vert k\rangle\langle k\vert$. Since $\tr[K]=1$, there exists a smallest natural number $n_0$ such that $\tr[P_{n_0} KP_{n_0}]\neq 0$. For each $n\geq n_0$, define the truncated operator $K_n =\frac{1}{\tr[P_n KP_n]} P_n KP_n$, where the normalization assures that it is a positive operator of unit trace. According to lemma \ref{infolemma}, each observable $\mathsf{E}^{K_n}$ is informationally complete. It is a well known fact that the sequence $(K_n)_{n\in\N}$ converges to $K$ in the trace norm. For each state $\rho$ and $X\in\h B(\C)$ we then have
\begin{eqnarray*}
\tr[ \rho \mathsf{E}^{K_n}(X)] &=& \int_X \tr[\rho D(z) K_n D(z)^*] \frac{d^2z}{\pi} =\int_X \tr[K_n D(z)^* \rho D(z)] \frac{d^2z}{\pi}\\
&=&\int_{-X} \tr[K_nD(z) \rho D(z)^*] \frac{d^2z}{\pi} =\tr[ K_n \mathsf{E}^\rho (-X)]
\end{eqnarray*}
and similarily for $K$. This then implies that
$$
\big\vert \tr[\rho \mathsf{E}^{K_n} (X)] -\tr[\rho \mathsf{E}^K (X)]\big\vert  =\big\vert \tr[K_n\mathsf{E}^\rho (-X)] -\tr[K\mathsf{E}^\rho (-X)]\big\vert \leq\Vert K_n-K\Vert_1 \Vert \mathsf{E}^\rho (-X)\Vert\rightarrow 0,
$$
as $n\rightarrow\infty$. In this way, the measurement of $\mathsf{E}^K$ is obtained as a limit of measurements of informationally complete observables. In particular, the measurement of an informationally {\em incomplete} observable can be obtained as such a limit.

\

\

\section{Matrix inversion results}\label{matrix}
In this section we prove the relevant results concerning the inversion of infinite matrices. The first result shows that any infinite upper triangular matrix with nonzero diagonal elements has a formal inverse.

First of all, notice that the product of two or more upper triangular matrices is always a well defined upper triangular matrix, in the sense that the matrix elements of the product matrix are well defined finite sums. To clarify this, consider the matrices  $A=(a_{mn})_{m,n\in\N}$ and $B=(b_{mn})_{m,n\in\N}$, for which $a_{mn}=0=b_{mn}$ for $n<m$. Now the matrix elements of the product matrix are
$$
(AB)_{m,m+l} =\sum_{k=0}^\infty a_{m,k}b_{k,m+l} =\sum_{k=m}^{m+l} a_{m,k}b_{k,m+l}
$$
for all $l\in\N$, and $(AB)_{m,n}=0$ for $n<m$. Similarly, any finite product of upper triangular matrices is well defined. If $C$ is a strictly upper triangular matrix, that is, the diagonal elements are zeros, then for each $m,l\in\N$ we have $(C^k)_{m,m+l}=0$ when $k>l$. In this way, the infinite series 
$$
\sum_{k=0}^\infty C^k
$$
is well defined in the sense that 
$$
\left(\sum_{k=0}^\infty C^k\right)_{m,m+l} =\sum_{k=0}^\infty  (C^k)_{m,m+l} =\sum_{k=0}^l  (C^k)_{m,m+l},
$$ 
that is, the series reduces to a finite sum for each $m,l\in\N$.

If $A$ is an upper triangular matrix with unit diagonal elements, then the matrix $(I-A)$ is stricly upper triangular. Thus, the series 
$$
\sum_{k=0}^\infty (I-A)^k
$$
is well defined in the above sense. The following lemma shows, that such a series is actually the formal inverse of $A$.

\begin{lemma}\label{ylakolmiomatriisi}
Let $A=(a_{mn})_{m,n\in\N}$ be an upper triangular infinite matrix with unit diagonal, that is, $a_{mn}=0$ for $n<m$, and $a_{mm}=1$ for all $m\in\N$, and let $B=(b_{mn})_{m,n\in\N}$ be a matrix for which $b_{mn}=0$ for $n<m$ and $b_{m,m+l}= \sum_{k=0}^l [(I-A)^k]_{m,m+l}$ for all $l\in\N$. Then $A$ and $B$ are formal inverses of each other, that is $(AB)_{mn}=\delta_{mn}=(BA)_{mn}$.
\end{lemma}
\begin{proof} 
First notice that for $n<m$ we have trivially $(AB)_{mn}=0=(BA)_{mn}$ since they involve empty sums. The case of the diagonal elements is also clear since for example $(AB)_{mm}=a_{mm}b_{mm}=1$. Suppose now that $n=m+l$, where $l>0$. Define the 
matrices $\tilde{A}:= (a_{ij})_{i,j=0}^{m+l}$ and $\tilde{B}:=(b_{ij})_{i,j=0}^{m+l}$ as finite cut-offs of the corresponding infinite matrices. Now
$$
(AB)_{m,m+l} =\sum_{k=m}^{m+l} a_{mk} b_{k,m+l}  =(\tilde{A}\tilde{B})_{m,m+l},
$$
so it is sufficient to prove the claim for finite matrices. Clearly
$$
\tilde{B} =\sum_{k=0}^{m+l}(I-\tilde{A})^k,
$$
since $(I-\tilde{A})^k =0$ when $k\geq m+l+1$. Thus
\begin{eqnarray*}
\tilde{A}\tilde{B} &=&(I-(I-\tilde{A}))\tilde{B} =\sum_{k=0}^{m+l} (I-\tilde{A})^k -\sum_{k=0}^{m+l-1}(I-\tilde{A})^{k+1}\\
&=& \sum_{k=0}^{m+l} (I-\tilde{A})^k -\sum_{k=0}^{m+l}(I-\tilde{A})^k +I=I,
\end{eqnarray*}
and hence
$$
(AB)_{m,m+l}=(\tilde{A}\tilde{B})_{m,m+l} =0
$$
for all $m\in\N$ and $l>0$. In a similar fashion one proves that $(BA)_{m,m+l} =0$ for all $m\in\N$, $l>0$.
\end{proof}

As an immediate consequence, we find the inverse of an upper triangular matrix with nonzero diagonal elements.
\begin{corollary}\label{seuraus}
Let $A=(a_{mn})_{m,n\in\N}$ be an upper triangular matrix with nonzero diagonal elements, and $U=(u_{mn})_{m,n\in\N}$ a diagonal matrix with $u_{mm}=a_{mm}^{-1}$. Then $A$ has a formal inverse $B=(b_{mn})_{m,n\in\N}$ such that $b_{mn}=0$ when $n<m$ and 
$$
b_{m,m+l} = \frac{1}{a_{m+l,m+l}} \sum_{k=0}^l [(I-UA)^k]_{m,m+l} =\frac{1}{a_{m,m}} \sum_{k=0}^l [(I-AU)^k]_{m,m+l}
$$ 
for all $m,l\in\N$.
\end{corollary}
\begin{proof} First notice that the problem again reduces to the case of finite matrices. Now $UA$ and $AU$ are upper triangular matrices with unit diagonals, so taking suitable cut-offs of these, the claim follows from elementary calculations.
\end{proof}

Consider now a finite sequence $(c_n)_{n=0}^k\subset \C$ in the above case. Define the sequence $(d_m)_{m\in\N}$ via 
$$
d_m = \sum_{n=0}^\infty a_{mn}c_n = \sum_{n=m}^k a_{mn}c_n.
$$ 
Since $d_m=0$ for $m>k$, the sequence is actually finite. Define $(c'_n)_{n\in\N}$ via 
$$
c'_n =\sum_{m=0}^\infty b_{nm}d_m =\sum_{m=n}^k b_{nm}d_m.
$$
Again, $c'_n =0$ for $n>k$, and inserting $d_m$ into the above equation gives us for $n\leq k$
$$c'_n =\sum_{m=n}^k \sum_{n'=m}^k b_{nm}a_{mn'} c_{n'} = \sum_{m=0}^k \sum_{n'=0}^k b_{nm}a_{mn'} c_{n'}
=\sum_{n'=0}^k \left(\sum_{m=0}^k b_{nm}a_{mn'}\right) c_{n'}=\sum_{n'=0}^k \delta_{nn'}c'_n =c_n
$$
since $n'\leq k$. This then implies that when restricted to the vector space of finite sequences, the linear mappings corresponding to the matrix and its formal inverse are inverse mappings of each other.

The second lemma deals with a special case of an upper triangular matrix, namely, one that is also an infinite dimensional Toeplitz matrix. That is, for all $l\in\N$, the $l$th diagonal elements $a_{m,m+l}$, $m\in\N$, do not depend on $m$. It turns out that the formal inverse $(b_{mn})_{m,n\in\N}$ is also an upper triangular Toeplitz matrix. In this case we also find a sufficient condition for inverting the relation
$$
d_m =\sum_{n=0}^\infty a_{mn}c_n,
$$
where $(c_n)_{n\in\N}$ is an infinite sequence, as 
$$
c_n = \sum_{m=0}^\infty b_{nm}d_m.
$$

\begin{lemma}\label{matriisilemma}
Let $l\in \N$, $l\geq 1$, $a_0,a_1,\ldots, a_l\in \C$, $a_0\neq 0$, and define the matrix $A=(a_{sn})_{s,n\in\N}$ for which $a_{sn} = a_{n-s}$, when $s\leq n\leq s+l$, and $a_{sn}= 0$ otherwise. Let $B=(b_{ns})_{n,s\in\N}$ be the formal inverse of $A$.
\begin{itemize}
\item[(a)] There exist a unique sequence $(b_u)_{u\in\N}\subset \C$ such  that $b_{ns}=b_{s-n}$ when $s\ge n$.
\item[(b)] Let $(c_n)_{n\in\N}\subset\C$ and define the sequence $(d_s)_{s\in\N}$ via $d_s =\sum_{n=0}^\infty a_{sn}c_n$. Suppose that for a given $n\in\N$, the condition $\lim_{m\rightarrow\infty} a_{k-n} b_{m-k} c_m =0 $ is satisfied for $k=n+1,\ldots, n+l$. Then 
$$
c_n =\sum_{s=0}^\infty b_{ns}d_s.
$$
\end{itemize}
\end{lemma}
\begin{proof} 
To prove $(a)$, we are going to show that the matrix elements $b_{n,n+k}$, $n,k\in\N$, do not depend on $n$. According to Corollary \ref{seuraus}, we have $b_{n,n+k}=\frac{1}{a_0} \sum_{u=0}^\infty [(I-UA)^u]_{n,n+k}$ for $k\geq 0$, and $b_{ns} =0$ otherwise. First note that $b_{nn} = \frac{1}{a_0}$ for all $n\in\N$. Suppose now that $k\geq 1$. Since in this case we have simply $UA =\frac{1}{a_0} A$, we get $(I-UA)_{ns} =-\frac{a_{s-n}}{a_0}$ for $s>n$, and $(I-UA)_{ns}=0$ otherwise. A direct calculation now gives us
\begin{eqnarray*}
[(I-UA)^u]_{n,n+k} &=&\sum_{t_1 =n+1}^\infty \sum_{t_2 =t_1+1}^\infty \cdots \sum_{t_{u-1} =t_{u-2}+1}^\infty (I-UA)_{nt_1} (I-UA)_{t_1t_2}\cdots (I-UA)_{t_{u-1},n+k}\\
&=&\left(-\frac{1}{a_0}\right)^u \sum_{t_1 =n+1}^{n+u} \sum_{t_2 =t_1+1}^{t_1+u} \cdots \sum_{t_{u-1} =t_{u-2}+1}^{t_{u-2} +u} a_{t_1-n}a_{t_2 -t_1}\cdots a_{n+k -t_{u-1}}
\end{eqnarray*}
for $u\geq 1$. After suitable changes in the summation indeces, we obtain
\begin{eqnarray*}
b_{n,n+k} &=&\frac{1}{a_0} +\sum_{u=0}^k \left(-\frac{1}{a_0}\right)^u \sum_{t_1 =n+1}^{n+u} \sum_{t_2 =t_1+1}^{t_1+u} \cdots \sum_{t_{u-1} =t_{u-2}+1}^{t_{u-2} +u} a_{t_1-n}a_{t_2 -t_1}\cdots a_{n+k -t_{u-1}}\\
&=&\frac{1}{a_0} +\sum_{u=0}^k \left(-\frac{1}{a_0}\right)^u \sum_{t_1 =1}^{u} \sum_{t_2 =t_1+1}^{t_1+u} \cdots \sum_{t_{u-1} =t_{u-2}+1}^{t_{u-2} +u} a_{t_1}a_{t_2 -t_1}\cdots a_{k -t_{u-1}},
\end{eqnarray*}
which goes to show that $b_{n,n+k}$ does not depend on $n$. Consequently, the sequence $(b_l)_{l\in\N}$, $b_l =b_{0l}$ is of the desired form. In addition, it is clearly unique.

To prove (b), we first deal with the case $n=0$. Consider the partial sum $S_k:=\sum_{s=0}^k b_{0s}d_s$, for $k\geq l-1$ (Recall the assumption $l\geq 1$.). We put in the expression
$$
d_s=\sum_{n'=0}^{\infty}a_{sn'}c_{n'}=\sum_{n'=s}^{s+l}a_{sn'}c_{n'}=\sum_{n'=0}^{k+l}a_{sn'}c_{n'}, \ \ k\geq s,$$ to get
$$
S_k = \sum_{s=0}^k b_{0s}\sum_{n'=0}^{k+l} a_{sn'}c_{n'}= 
\sum_{n'=0}^{k+l}\left(\sum_{s=0}^k b_{0s} a_{sn'}\right)c_{n'}.
$$
According to (a), the sum in parenthesis equals $\delta_{0n'}$, provided that the summation covers the full range of nonzero $a_{sn'}$:s. This happens exactly when $k\geq n'$. Thus we get
\begin{eqnarray*}
S_k &=& c_0 + \sum_{n'=k+1}^{k+l} \left(\sum_{s=0}^k b_{0s} a_{sn'}\right)c_{n'}=
c_0 + \sum_{n'=k+1}^{k+l} \left(\sum_{s=n'-l}^k b_s a_{n'-s}\right)c_{n'}\\
&=& c_0 + \sum_{n'=1}^{l} \left(\sum_{s=n'+k-l}^k b_s a_{n'+k-s}\right)c_{n'+k}
= c_0 + \sum_{n'=1}^{l} \left(\sum_{s=n'}^{l} b_{n'+k-s}a_{s}\right)c_{n'+k}
\end{eqnarray*}
where the third equality is obtained by substituting $n'\mapsto n'+k$ in the outer sum, and the fourth equality by substituting $s\mapsto n'+k-s$ in the inner sum. Suppose now that the limit condition holds for $n=0$. Then
$$
0 = \lim_{k\rightarrow \infty}\sum_{n'=1}^{l} \left(\sum_{s=n'}^{l} b_{k-s}a_{s}\right)c_{k}
= \lim_{k\rightarrow \infty}\sum_{n'=1}^{l} \left(\sum_{s=n'}^{l} b_{n'+k-s}a_{s}\right)c_{n'+k}
$$
proving that $\lim_{k\rightarrow \infty} S_k=c_0$.

Now fix an $n\in \N$ and define a translated sequence $\tilde{c}_{n'}=c_{n'+n}$, $n'\in \N$, with
$$
\tilde{d}_s := \sum_{n'=0}^\infty a_{sn'}\tilde{c}_{n'} = \sum_{n'=s}^{s+l}a_{n'-s}c_{n+n'} =
\sum_{n'=s+n}^{s+n+l}a_{n'-(s+n)}c_{n'} =d_{s+n}, \ \ s\in \N.
$$
Hence, we have the convergence
$$
c_n=\sum_{s=0}^\infty b_{ns}d_s = \sum_{s=n}^\infty b_{ns}d_s = \sum_{s=0}^\infty b_{n,s+n}d_{s+n} = \sum_{s=0}^\infty b_{0s}d_{s+n}
$$
exactly when
$$
\tilde{c}_0=\sum_{s=0}^\infty b_{0s}\tilde{d}_s
$$
which, according to the result just obtained, happens if and only if
$\lim_{m\rightarrow\infty} a_k b_{m-k}\tilde{c}_m =0$, $k=1,\ldots, l$. But this is equivalent to the claimed limit condition, and the proof is complete.
\end{proof}

\begin{remark}\rm 
\begin{itemize}
\item[(a)] According to the proof, a necessary and sufficient condition for the convergence of the series  $c_n=\sum_{s=0}^\infty b_{ns}d_s$ is that the remainder
$$R^n_k := \sum_{n'=1}^l\sum_{s=n'}^l b_{n'+k-s}a_s\, c_{n'+n+k}$$ goes to zero in the limit
$k\rightarrow \infty$. This is not equivalent to the limit condition of lemma \ref{matriisilemma} in general.
\item[(b)] The limit condition of lemma \ref{matriisilemma} cannot be relaxed. Indeed, at least in the case $l=1$, it is also necessary for the convergence of the series, if we  assume $a_1\neq 0$. This is apparent, since the remainder term $R_k^n$ contains only one term then. Another example is  given by $l=2$, $a_1=0$, and $a_2\neq 0$. In this case,
$b_{2s}= (-\frac{a_2}{a_0})^s\frac{1}{a_0}$, and $b_{2s+1}=0$. Hence the remainder term is
$R_k^n= a_2b_{k-1}c_{n+k+1}$ for odd $k$, and $R_k^n=a_2b_kc_{n+k+2}$
for even $k$. Thus, the necessary and sufficient condition for the convergence for the $c_n$ series is $\lim_{k\rightarrow \infty} a_2b_{2k}c_{2k+n+2}=0$. This is just the same as the limit condition 
for all sequences $(c_n)_{n\in\N}$, since $a_1=0$ implies that
$\lim_{k\rightarrow \infty} a_{1}b_{2k}c_{2k+n+1}=0$ trivially.
\end{itemize}
\end{remark}

\section{Phase space observables generated by the number states}\label{suure}

For each $s\in\N$, let $G^{\vert s\rangle}:[0,\infty)\times [0,2\pi)\to \h L(\hil)$ be the operator density associated with the phase space observable generated by the number state $\vert s\rangle\langle s\vert$, i.e. $G^{\vert s\rangle}(r,\theta)=D(re^{i\theta})|s\rangle \langle s|D(re^{i\theta})^*$. For any state $\rho$, let $G^{\vert s\rangle}_\rho(r,\theta):=\tr[\rho G^{\vert s\rangle}(r,\theta)]$ be the corresponding probability density\footnote{The function $(s,z)\mapsto \omega(s,z):=\tr[\rho G^{\vert s\rangle}(z)]$ is also known as the photon number tomogram \cite{tomogram1, tomogram2}, that is, the tomogram is identified with the collection of probability densities $G^{\vert s\rangle}_\rho$, $s\in\N$.} . The informational completeness of the corresponding observable
$$
\h B(\C)\ni Z\mapsto \mathsf{E}^{\vert s\rangle} (Z) =\int_Z D(z)\vert s\rangle\langle s\vert D(z)^* \frac{d^2z}{\pi}\in \h L(\hil)
$$
follows directly from lemma \ref{infolemma}, and thus the reconstruction  of the state is, in principle, possible from the measured distribution.

The matrix elements of the operator density $G^{\vert s\rangle}$ with respect to the number basis are
$$
\langle n\vert G^{\vert s\rangle}(r,\theta) \vert m\rangle =e^{i\theta(n-m)} f^s_{nm}(r),
$$
where 
$$
f^s_{nm}(r) :=\langle n\vert D(r)\vert s\rangle\langle s\vert D(r)^*\vert m\rangle.
$$
Thus, the probability density $G^{\vert s\rangle}_\rho (r,\theta) =\tr[\rho G^{\vert s\rangle} (r,\theta)]$ can be written as
$$
G^{\vert s\rangle}_\rho(r,\theta) =\sum_{m,n=0}^\infty \rho_{mn} \langle n\vert G^{\vert s\rangle} (r,\theta)\vert m\rangle = \sum_{m,n=0}^\infty \rho_{mn} e^{i\theta(n-m)} f^s_{nm}(r).
$$
Using equation \eqref{Dmatriisi}, the explicit form of the functions $f^s_{n,m}$ can be written as
\begin{eqnarray}\label{f}
f^s_{nm}(r) &=& (-1)^{\max\{ 0,s-n\} +\max\{0,s-m\}} \sqrt{\frac{\min \{n,s\}!\min \{m,s\}!}{\max\{n,s\}!\max\{m,s\}!}} \nonumber\\
&& \times e^{-r^2} r^{\vert s-n\vert+\vert s-m\vert} L^{\vert s-n\vert}_{\min \{n,s\}} (r^2)L^{\vert s-m\vert}_{\min \{m,s\}} (r^2).
\end{eqnarray}

The mapping $\theta\mapsto G^{\vert s\rangle} (r,\theta)$ is weakly continuous for each $r\in[0,\infty)$, and $\Vert G^{\vert s\rangle} (r,\theta)\Vert =1$ for all $r\in[0,\infty)$, $\theta\in[0,2\pi)$, so the operator 
$$
G^{\vert s\rangle}_l (r) := \frac{1}{2\pi} \int_0^{2\pi} e^{il\theta} G^{\vert s\rangle} (r,\theta)\, d\theta
$$
is well-defined as a weak integral. In addition, we have
$$
G^{\vert s\rangle}_{\rho,l} (r) :=\tr[\rho G^{\vert s\rangle}_l (r)] =\frac{1}{2\pi} \int_0^{2\pi} e^{il\theta}G^{\vert s\rangle}_\rho (r,\theta)\, d\theta,
$$
for all states $\rho$. A simple calculation gives us
$$
G^{\vert s\rangle}_{\rho,l} (r) =\sum_{n=0}^\infty \rho_{n+l,n}\langle n\vert D(r)\vert s\rangle\langle s \vert D(r)^* \vert n+l\rangle,
$$
for all $r\in [0,\infty)$.

The probability distributions $G^{\vert s\rangle}_\rho$, also known as displaced photon distributions, are closely related to the $\la$-parametrized phase space quasiprobability distributions, first presented by Cahill and Glauber \cite{Cahill1969, Glauber1969}. To clarify this, let us recall the definition of these distributions. For each $\la\in\C$, $\vert\la\vert <1$, define the operator density  $W^\la :[0,\infty)\times [0,2\pi)\rightarrow \h L(\hil)$ by
$$
W^\la (r,\theta) := (1-\la)\sum_{k=0}^\infty \la^k D(re^{i\theta})\vert k\rangle\langle k\vert D(re^{i\theta})^\ast,
$$
and the corresponding probability density $W^\la_\rho$ by $W^\la_\rho(r,\theta) =\tr [\rho W^\la (r,\theta)]$. It is clear from these definitions, that indeed, one has 
$$
W^\la_\rho (r,\theta) =(1-\la) \sum_{k=0}^\infty \la^k  G^{\vert k\rangle}_\rho(r,\theta).
$$

To obtain the displaced photon distributions from the $\la$-distribution, we first note that 
$$
\vert G^{\vert k\rangle}_\rho (r,\theta)\vert =\vert\langle k\vert D(re^{i\theta})^* \rho D(re^{i\theta})\vert k\rangle\vert \leq \Vert \rho\Vert =\Vert \rho\Vert_1\leq 1,
$$
for all $r\in[0,\infty)$, $\theta\in [0, 2\pi)$, which follows from the Cauchy-Schwarz inequality. This then implies that $(1-\la)^{-1} W^\la_\rho (r,\theta)= \sum_{k=0}^\infty \la^k G^{\vert k\rangle}_\rho (r,\theta)$ is a power series with respect to $\la$, converging absolutely for all $\la\in\C$, $\vert\la\vert <1$, suggesting that the series can be differentiated around the origin term by term. 
A direct calculation now gives us
\begin{equation}\label{lambdaderivaatta}
\frac{1}{s!}\frac{\partial^s}{\partial\la^s} \big( (1-\la)^{-1}  W^\la_\rho (r,\theta)\big)\bigg\vert_{\la=0} = \frac{1}{s!} \sum_{k=0}^\infty \frac{\partial^s\la^k}{\partial\la^s}\bigg\vert_{\la=0} G^{\vert k\rangle}_\rho (r,\theta)  =G^{\vert s\rangle}_\rho (r,\theta),
\end{equation}
since $\frac{\partial^s\la^k}{\partial\la^s}\big\vert_{\la=0} =s!\delta_{sk}$.

These of course give us, at least in principle, the possibility of constructing either of the distributions from the other. In a recent paper \cite{KiPeSc}, rigorous proofs for two reconstruction formulas for the $\la$-distributions were given. In view of this, the knowledge of all of the distributions $G^{\vert s\rangle}_\rho$, $s\in\N$, allows state reconstruction via a detour.

\subsection{Measuring the displaced photon distributions}
We will now review the possibility of measuring the $G^{\vert s\rangle}$-distributions with an eight-port homodyne detection scheme. For a basic reference concerning the setup, see e.g. \cite{Leonhardt}. In \cite{eightport} a rigorous proof was given for the fact that with this scheme, any covariant phase space observable can be obtained as a high amplitude limit. The detector consists of two pairs of photon detectors and the amplitude-scaled photon differences $D_1$ and $D_2$ are measured. Four input modes are involved; the signal mode, a vacuum mode, a local oscillator in a coherent state, and a parameter mode which defines the observable to be measured. If the parameter mode is in a state $S$, then the phase space observable $E^{CSC^{-1}}$, where $C$ is the conjugation map $\psi\mapsto (x\mapsto \overline{\psi(x)})$, can be obtained as the high amplitude limit (see \cite{eightport}).

The first obvious way to measure the $G^{\vert s\rangle}$-distributions is the direct measurement in the sense of the above limit. However, this requires ideal detectors, and the parameter field needs to be prepared in a number state $\vert s\rangle\langle s\vert$. The preparation of the number state is highly untrivial and is by itself an active area of research. Several theoretical models, mostly in the context of cavity quantum electrodynamics, for the preparation of an arbitrary number state have been proposed (see e.g. \cite{Geremia2006, Brown2003, Brattke2001, Bueno2008}). Even though this gives a theoretical method for measuring the distributions, it is not a practical one since the preparation procedures work only for small photon numbers. To avoid the problem of number state preparation, we consider an alternative point of view.

Consider the measurement of the $Q$-function of the electromagnetic field by means of the above experimental setup. In this case the parameter field is in the vacuum state $\vert 0\rangle\langle 0\vert$. If the detectors are non-ideal, with a detection efficiency $\eta$ each, the measured distribution is actually the $\la$-parameterized distribution, with $\la = 1-\eta$ \cite{Leonhardt1993, DAriano1995}. Suppose now that the detector efficiencies are close to unity, that is $\la \approx 0$. Then, by adding suitable beam-splitters into the measurement scheme, one is able to measure the distributions corresponding to the parameter $\la'$, for which $\la'\geq \la$. An equivalent scheme would be one where the detector efficiencies could be adjusted. Proceeding in this manner, one obtains a function $\la\mapsto W^\la_\rho$. In an ideal situation where $\eta =1$ one could thus differentiate this $s$ times with respect to $\la$, and obtain the $G^{\vert s\rangle}$-distribution according to equation \eqref{lambdaderivaatta}. Even in the nonideal case, one can obtain some kind of an approximation for the $G^{\vert s\rangle}$-distributions, provided that the $\la$-dependence of $W^\la_\rho$ is regular enough to allow an extrapolation to the values close to the origin.

\subsection{Reconstruction from the set $\{ G_\rho^{\vert s\rangle} \vert s\in\N\}$ of distributions}
If one has knowledge of all of the distribution $G^{\vert s\rangle}_\rho$, $s\in\N$, recovering the diagonal elements of the density matrix is a trivial task. Indeed, by definition one has
$$
G^{\vert s\rangle}_\rho (0) = \tr [\rho D(0)\vert s\rangle\langle s\vert D(0)^*] =\langle s\vert \rho\vert s\rangle =\rho_{ss},
$$
suggesting that in order to reconstruct the diagonal elements of the state matrix, one needs to measure the observable $\mathsf{E}^{\vert s\rangle}$ around the origin for all $s\in\N$. The reconstruction of the off-diagonal elements is a more complicated matter.

Let $l\in\N$, $l\geq 1$, so that 
$$
G^{\vert s\rangle}_{\rho,l} (r) =\sum_{n=0}^\infty \rho_{n+l,n} f^s_{n,n+l}(r),
$$
where the functions $f^s_{n,n+l}$ were defined in equation \eqref{f}. Define a function $g_l:(0,\infty)\rightarrow \C$ via $g_l (r)=e^{r^2} r^{-l}$. Suppose that $l\leq s$. Then the limit $T^l_{sn}:=\lim_{r\rightarrow 0} (g_lf_{n,n+l}^s)(r)$ exists, because $|s-n|+|s-(n+l)|\geq l$  for any $n\in \N$, $s\geq l$,
and can easily be computed using the fact that
$L_m^\alpha(0)= \binom{m+\alpha}{m}$; the result is
$$
T^l_{sn}=\begin{cases} 0, & n<s-l;\\
(-1)^{s-n}\sqrt{\frac{(n+l)!}{n!}}\frac{1}{(s-n)!(n+l-s)!}, & s-l\leq n\leq s;\\
0, & n> s,\end{cases}
$$
In addition, assuming $n\geq s$ and $r\in(0,1)$, and using the fact that $|L_s^{n-s}(r^2)|\leq \binom n s e^{\frac 12r^2}$ \cite[p. 786, 22.14.12]{Abramowitz} we get
$$
|g_l(r)f_{n,n+l}^s(r)|\leq \frac{e}{s!}\sqrt{\frac{n^{s}}{(n-s)!}\frac{(n+l)^{s}}{(n+l-s)!}}
$$
which goes to zero, as $n\rightarrow\infty$. This implies that
$\sup_{n\in \N, r\in (0,1)} |g_l(r)f_{n,n+l}^s(r)|<\infty$. Since $\sum_{n=0}^\infty \vert \rho_{n+l,n}\vert \leq 1$, it follows that the series $\sum_{n=0}^\infty \rho_{n+l,n} g_l (r) f_{n,n+l} (r)$ converges absolutely and uniformly on the interval $(0,1)$. Thus, the limit 
$$
\lim_{r\rightarrow 0} g_l (r) G^{\vert s\rangle}_{\rho,l} (r) = \lim_{r\rightarrow 0} \sum_{n=0}^\infty \rho_{n+l,n} g_l (r) f^s_{n,n+l}(r)
$$ 
may be taken termwise. This gives us the infinite matrix identity
$$
d^l_s:=\lim_{r\rightarrow 0}g_l(r)G_{\rho,l}^{\vert s\rangle}(r)=\sum_{n=0}^\infty T^l_{sn}\rho_{n+l,n},
$$
which, in this case, holds for all states $\rho$. Inserting the explicit form of $T^l_{sn}$ we obtain
\begin{eqnarray}\label{raja}
d^l_s &=&\sum_{n=s-l}^s (-1)^{s-n} \sqrt{\frac{(n+l)!}{n!}} \frac{1}{(s-n)!(n+l-s)!} \rho_{n+l,n} \nonumber\\
&=&\sum_{n'=s}^{s+l} (-1)^{s+l-n'} \sqrt{\frac{n'!}{(n'-l)!}} \frac{1}{(s+l-n')!(n'-s)!} \rho_{n',n'-l}.
\end{eqnarray}

Defining $c_n^l:=\frac{(-1)^l}{l!} \sqrt{\frac{n!}{(n-l)!}}  \rho_{n,n-l}$ for $n\geq l$ and $c_n^l=0$ otherwise, and $a_{sn}^l:= (-1)^{n-s} \binom{l}{n-s}$, we can write equation \eqref{raja} as
$$
d_s^l =\sum_{n=0}^\infty a_{sn}^l c_n^l,
$$
since $s\geq l$. The infinite matrix $(a_{sn}^l)_{s,n\in\N}$ is now of the type considered in lemma \ref{matriisilemma} with $a_u^l =(-1)^{u} \binom{l}{u}$. Consider now the sequence $(b_u^l)_{u\in\N}$ with $b_u^l =\binom{u+l-1}{l-1}$, and the infinite matrix $(b_{ns}^l)_{n,s\in\N}$, $b_{ns}^l =b_{s-n}^l$. This is an upper triangular matrix, and we have
$$
\sum_{n=0}^\infty a^l_{sn} b^l_{ns} = \sum_{s=0}^\infty b^l_{ns} a^l_{sn}  = a^l_{nn} b^l_{nn}= \binom{l}{0} \binom{l-1}{l-1} =1.
$$
To calculate the off-diagonal elements of the product matrices, let $k\geq 1$. Using formula (5) on page 8 of \cite{Riordan}, we find that
\begin{eqnarray*}
\sum_{n=0}^\infty a^l_{sn} b^l_{n,s+k} &=&\sum_{n=s}^\infty (-1)^{n-s} \binom{l}{n-s} \binom{s+k -n+l-1}{l-1} \\
&=& \sum_{n'=0}^\infty (-1)^{n'} \binom{l}{n'} \binom{k -n'+l-1}{k-n'} =\binom{k-1}{k} =0.
\end{eqnarray*}
For the other case we use formula (5d) on page 10 of \cite{Riordan} to obtain
\begin{eqnarray*}
\sum_{s=0}^\infty b^l_{ns} a^l_{s,n+k} &=& \sum_{s=n}^\infty (-1)^{n+k-s} \binom{l}{n+k-s} \binom{s -n+l-1}{l-1} \\
&=& \sum_{s'=0}^\infty (-1)^{k+s'} \binom{l}{l-k+s'} \binom{s'+l-1}{s'} =(-1)^{l+1-k}\binom{k-1}{k} =0.
\end{eqnarray*}
Since the lower diagonal elements of the product matrices are zero by definition, we find that $(a_{sn}^l)_{s,n\in\N}$ and $(b_{ns}^l)_{n,s\in\N}$ are formal inverses of each other.

Suppose now that 
\begin{equation}\label{ehto}
\lim_{m\rightarrow\infty} m^{\frac{3}{2}l-1} \rho_{m,m-l} =0
\end{equation}
for all $l\in\N$. Then the limit condition of lemma \ref{matriisilemma} is satisfied for each $n\in\N$, since
\begin{eqnarray*}
\vert b_{m-k}^l c_m^l\vert &=& \frac{1}{l!} \sqrt{\frac{m!}{(m-l)!}} \frac{(m+l-k-1)!}{(l-1)!(m-k)!} \vert \rho_{m,m-l}\vert \\
&\leq & \frac{1}{l!(l-1)!} m^{\frac{l}{2}} (m+l-k-1)^{l-1}\vert \rho_{m,m-l}\vert\\
&\leq&  2^{l-1}m^{\frac{3}{2}l-1} \vert\rho_{m,m-l}\vert
\end{eqnarray*}
for $m\geq l$. Under this condition, we then have the convergence
$$
c_n^l =\sum_{s=0}^\infty b_{ns}^l d_s^l=\sum_{s=n}^\infty b_{ns}^l d_s^l,
$$
or equivalently
$$
c_{n+l}^l =\sum_{s=n+l}^\infty b_{n+l,s}^l d_s^l,
$$
which gives us the reconstruction formula
\begin{equation}\label{rekonstruktio1}
\rho_{n+l,n} =(-1)^l l!\sqrt{\frac{n!}{(n+l)!}} \sum_{s=n+l}^\infty \binom{s-n-1}{l-1} d^l_s,
\end{equation}
where $d^l_s$ is a quantity which can be calculated directly from the measurement statistics.

\begin{remark}\rm Notice that the condition \eqref{ehto} for a given $l\in\N$ is a sufficient condition for the reconstruction of the $l$th diagonal of the density matrix. For $l=0$, for example, the reconstruction formula works for all states $\rho$. However, in the general case, the validity of the formula depends on the state in question. To illustrate this fact, consider the vector state $\psi =\frac{\pi}{\sqrt{6}}\, \sum_{n=1}^\infty \frac{1}{n} \vert n\rangle$. Condition \eqref{ehto} now states that we should have
$$
\lim_{n\rightarrow\infty}  \frac{m^{\frac{3}{2}l-1}}{m(m-l)} =0,
$$
which is clearly not true for $l\geq 2$. It is easy to check that even the weaker condition, namely the limit condition of lemma \ref{matriisilemma}, is unsatisfied. This then suggests that the reconstruction formula \eqref{rekonstruktio1} does not work for all states.
\end{remark}

\subsection{Reconstruction from a single distribution}
If we want to use a single distribution $G^{\vert s\rangle}_\rho$, the reconstruction formula becomes more complicated, and we were not able to satisfactorily solve the convergence issues in the case of an infinite density matrix. Consequently, we will assume in the sequel that the matrix is finite. This corresponds to the discussion of section \ref{matrix} concerning finite sequences.

The reconstruction makes use of the connection to the $\la$-parameterized distributions $W^\la_\rho$. It follows from equation \eqref{lambdaderivaatta}, that
$$
G^{\vert s\rangle}_{\rho,l} (r,\theta) = \frac{1}{s!}\frac{\partial^s}{\partial\la^s} \big( (1-\la)^{-1}  W^\la_{\rho,l} (r,\theta)\big)\bigg\vert_{\la=0}
$$
for all $l\in\N$. On the other hand, we have for all states $\rho$ and $l\in\N$ 
$$
W^\la_{\rho,l} (r):=\frac{1}{2\pi}\int_0^{2\pi}e^{il\theta} \tr[\rho W^\la (r,\theta)]\, d\theta =\sum_{n=0}^\infty \rho_{n+l,n} K^\la_{n,n+l} (r),
$$
where
\begin{equation}\label{K}
K^\la_{n,n+l} (r) = (1-\la)\sum_{k=0}^\infty \la^k \langle n\vert D(r)\vert k\rangle\langle k\vert D(r)^* \vert n+l\rangle.
\end{equation}
This series can again be differentiated termwise, and we get
$$
G^{\vert s\rangle}_{\rho,l} (r,\theta) = \sum_{n=0}^\infty \rho_{n+l,n} \frac{1}{s!}\frac{\partial^s}{\partial\la^s} \big( (1-\la)^{-1}K^\la_{n,n+l} (r)\big)\bigg\vert_{\la=0}.
$$
The explicit form of the functions $K_{n,n+l}$ is given by the formula of Cahill and Glauber \cite{Cahill1969}
\begin{eqnarray*}
K^\la_{n,n+l}(r) &=&
\sqrt{\frac{n!}{(n+l)!}}(1-\la)^{l+1}e^{-(1-\la)r^2}r^{l}\la^n L^{l}_n\big(
(2-\la-\la^{-1})r^2\big)\\
&=&\sqrt{n!(n+l)!} \sum_{u=0}^n \frac{(1-\la)^{2u+l+1} \la^{n-u} r^{2u+l}}{(n-u)!(l+u)! u!} e^{-(1-\la)r^2}.
\end{eqnarray*}
Before proceeding any further, we prove the following lemma.

\begin{lemma}\label{binomiallemma}
Let $k,\,p,\,q,\,s\in\N$ and $x\in\R$.
\begin{itemize}
\item[(a)]
$$
\frac{1}{k!}\frac{d^k (1-\la)^p\la^q}{d\la^k}\bigg|_{\la=0}=(-1)^{k+q}{p\choose k-q}
$$
which is 0 if and only if $k<q$ or $k>p+q$.
\item[(b)]
$$
\frac{1}{s!}\frac{d^s (1-\la)^p\la^q e^{\la x}}{d\la^s}\bigg|_{\la=0}=\sum_{k=q}^{\min\{s,\,p+q\}}\frac{(-1)^{k+q}}{(s-k)!}{p\choose k-q}x^{s-k}.
$$
which is 0 for all $x$ if and only if $s<q$.
\end{itemize}
\end{lemma}
\begin{proof}
By direct calculation we get
$$
\frac{d^k (1-\la)^p\la^q}{d\la^k}\bigg|_{\la=0}=\sum_{t=0}^k{k\choose t}\underbrace{\frac{d^t \la^q}{d\la^t}\bigg|_{\la=0}}_{=\,q!\delta_{q,t}}\frac{d^{k-t}(1-\la)^p}{d\la^{k-t}}\bigg|_{\la=0}
$$
from which (a) follows. Part (b) follows from (a) and the calculation
\begin{eqnarray*}
\frac{1}{s!}\frac{d^s (1-\la)^p\la^q e^{\la x}}{d\la^s}\bigg|_{\la=0}&=&
\frac{1}{s!}\sum_{k=0}^s{s\choose k}\frac{d^k (1-\la)^p\la^q}{d\la^k}\bigg|_{\la=0}
\frac{d^{s-k}e^{\la x}}{d\la^{s-k}}\bigg|_{\la=0}\\
&=&\sum_{k=0}^s\frac{(-1)^{k+q}}{(s-k)!}{p\choose k-q}x^{s-k}.
\end{eqnarray*}
\end{proof}

Now fix $s\in\N$ and denote $x=r^2$. We have two different cases depending on whether $l$ is even or odd. We will start with the even case.

\

\noindent\textbf{The even case.} Suppose that $l=2h$ for some  $h\in\N$. Then, by lemma \ref{binomiallemma} we get
\begin{eqnarray*}
\frac1{s!}\frac{\pa^s}{\pa \la^s}\big( (1-\la)^{-1}K^\la_{n,n+2h}(\sqrt{x})\big)\Big|_{\la=0}
&=&e^{-x}\sqrt{{n!}{(n+2h)!}}\sum_{u=\max\{0,n-s\}}^n\frac{1}{u!(n-u)!(u+2h)!}\times\\
&&\times
\sum_{k={n-u}}^{\min\{s,\,{u+2h}+n\}}\frac{(-1)^{k+{n-u}}}{(s-k)!}{{2(u+h)}\choose k-(n-u)}x^{u+h+s-k}. 
\end{eqnarray*}
For any $t\in\N$, define 
$$
H^s_{2h}(t,n):=
\frac{\pa^t}{\pa x^t}\frac{e^x}{s!}\frac{\pa^s}{\pa \la^s}\big( (1-\la)^{-1}K^\la_{n,n+2h}(\sqrt{x})\big)\Big|_{\la=0}\Big|_{x=0}
$$
so that
$$
\frac{\pa^t}{\pa x^t}{e^x}G^{\vert s\rangle}_{\rho,2h}(\sqrt{x})\Big|_{x=0}
=\sum_{n=0}^\infty \rho_{n+2h,n}H^s_{2h}(t,n).
$$
Now
\begin{eqnarray*}
H^s_{2h}(t,n)&=&
\sum_{u=\max\{0,n-s\}}^n\frac{\sqrt{{n!}{(n+2h)!}}}{u!(n-u)!(u+2h)!}
\sum_{k={n-u}}^{\min\{s,\,{u+2h}+n\}}\frac{(-1)^{k+{n-u}}}{(s-k)!}{{2(u+h)}\choose k-(n-u)}
\underbrace{\frac{\pa^t x^{u+h+s-k}}{\pa x^t}\Big|_{x=0}}_{=\;t!\,\delta_{t,u+h+s-k}} \\
&=&
\sum_{u=\max\{0,n-s\}}^n\frac{t!\sqrt{{n!}{(n+2h)!}}}{u!(n-u)!(u+2h)!}
\frac{(-1)^{s+h+t+n}}{(t-h-u)!}{{2(u+h)}\choose h+s-t-n+2u} \\
&=&\underbrace{
\frac{t!\sqrt{{n!}{(n+2h)!}}(-1)^{s+h+t+n}}{(h+t+n-s)!}}_{=\;0\text{ iff }h+t+n-s<0}
\underbrace{\sum_{u=\max\{0,n-s\}}^{\min\{n,t-h\}}\frac{1}{(n-u)!(u-n+s)!}
{2(u+h)\choose u}{s-n+u \choose t-h-u}}_{=\;0\text{ iff $t-h<0$ or $t-h<n-s$ or $n<t-h-s$}}.
\end{eqnarray*}
Since $H^s_{2h}(t,n)=0$ if $t<h$ we next assume that $t\ge h$ and get
\begin{equation}\label{eq10}
\frac{\pa^t}{\pa x^t}{e^x}G^{\vert s\rangle}_{\rho,2h}(\sqrt{x})\Big|_{x=0}
=\sum_{n=\max\{0,s-h-t,t-h-s\}}^{s-h+t} H^s_{2h}(t,n)\rho_{n+2h,n}.
\end{equation}

Let $t\ge s+h$ and denote $p=t-s-h$. We have 
$$
\frac{\pa^{p+s+h}}{\pa x^{p+s+h}}{e^x}G^{\vert s\rangle}_{\rho,2h}(\sqrt{x})\Big|_{x=0}
=\sum_{n=p}^{2s+p} H^s_{2h}(p+s+h,n)\rho_{n+2h,n}.
$$
Define an upper triangular matrix $(A^{s,2h}_{pn})_{p,n\in\N}$ by
$$
A^{s,2h}_{pn}:=H^s_{2h}(p+s+h,n),\hspace{1cm}n\ge p.
$$
According to Corollary \ref{seuraus}, it has an inverse matrix
$$
(B^{s,2h}_{np})_{n,p\in\N}.
$$
Using this, we get the state reconstruction formula 
\begin{equation}\label{rekonstruktio2a}
\rho_{n+2h,n} =\sum_{p=0}^\infty B^{s,2h}_{np} \frac{\pa^{p+s+h}}{\pa x^{p+s+h}} e^x G^{\vert s\rangle}_{\rho,2h} (\sqrt{x})\Big|_{x=0}.
\end{equation}

\

\noindent\textbf{The odd case.} If $l$ is odd, that is, $l=2h+1$ we get
\begin{eqnarray*}
&&\frac1{s!}\frac{\pa^s}{\pa \la^s}\big( (1-\la)^{-1}K^\la_{n,n+2h+1}(\sqrt{x})\big)\Big|_{\la=0}
=\sqrt{x}e^{-x}\sqrt{{n!}{(n+2h+1)!}}\\
&&\times\sum_{u=\max\{0,n-s\}}^n\frac{1}{u!(n-u)!(u+2h+1)!}\sum_{k={n-u}}^{\min\{s,\,u+2h+1+n\}}\frac{(-1)^{k+{n-u}}}{(s-k)!}{{2(u+h)+1}\choose k-(n-u)}x^{u+h+s-k}. 
\end{eqnarray*}
For all $t\in\N$, define 
$$
H^s_{2h+1}(t,n):=
\frac{\pa^t}{\pa x^t}\frac{\sqrt x\,e^x}{s!}\frac{\pa^s}{\pa \la^s}\big( (1-\la)^{-1}K^\la_{n,n+2h+1}(\sqrt{x})\big)\Big|_{\la=0}\Big|_{x=0}
$$
and calculate
\begin{eqnarray*}
H^s_{2h+1}(t,n)&=&
\underbrace{
\frac{t!\sqrt{{n!}{(n+2h+1)!}}(-1)^{s+h+t+n}}{(h+t+n-s+1)!}}_{=\;0\text{ iff }h+t+n-s+1<0}\times \\
&&\times
\underbrace{\sum_{u=\max\{0,n-s\}}^{\min\{n,t-h\}}\frac{1}{(n-u)!(u-n+s)!}
{2(u+h)+1\choose u}{s-n+u \choose t-h-u}}_{=\;0\text{ iff $t-h<0$ or $t-h<n-s$ or $n<t-h-s$}}.
\end{eqnarray*}
Since $H^s_{2h+1}(t,n)=0$ if $t<h$ we next assume that $t\ge h$ and get the infinite matrix identity
\begin{equation}\label{eq11}
\frac{\pa^t}{\pa x^t}{\sqrt x\,e^x}G^{\vert s\rangle}_{\rho,2h+1}(\sqrt{x})\Big|_{x=0}
=\sum_{n=\max\{0,s-h-t-1,t-h-s\}}^{s-h+t} H^s_{2h+1}(t,n)\rho_{n+2h+1,n}.
\end{equation}

Similarly to the even case, assume $t\ge s+h$ and denote $p=t-s-h$ to get
$$
\frac{\pa^{p+s+h}}{\pa x^{p+s+h}}{\sqrt x\,e^x}M^s_{\rho,2h+1}(\sqrt{x})\Big|_{x=0}
=\sum_{n=p}^{2s+p} H^s_{2h+1}(p+s+h,n)\rho_{n+2h+1,n}.
$$
Let $(B^{s,2h+1}_{np})_{n,p\in\N}$ be the inverse of an upper triangular matrix $(A^{s,2h+1}_{pn})_{p,n\in\N}$ 
with
$$
A^{s,2h+1}_{pn}:=H^s_{2h+1}(p+s+h,n),\hspace{1cm}n\ge p.
$$
Thus, we get the formula
\begin{equation}\label{rekonstruktio2b}
\rho_{n+2h+1,n} =\sum_{p=0}^\infty  B^{s,2h+1}_{np} \frac{\pa^{p+s+h}}{\pa x^{p+s+h}} \sqrt{x}e^x G^{\vert s\rangle}_{\rho,2h+1} (\sqrt{x})\Big|_{x=0}.
\end{equation}

\begin{example}\rm
As an illustrative example, we consider the simple case of the observable $\mathsf{E}^{\vert 1\rangle}$ generated by the first number state $\vert 1\rangle\langle 1\vert$, and the system in the state $\rho =\sum_{k=0}^2 \alpha_k \vert k\rangle\langle k\vert$, where $\alpha_k\geq0$ for $k=0,1,2$ and $\sum_{k=0}^2 \alpha_k=1$. The density matrix is now diagonal, so we use the reconstruction formula \eqref{rekonstruktio2a} for $s=1$, $h=0$ and $n=0,1,2$. We easily obtain the quantity
$$
e^x G^{\vert 1\rangle}_{\rho,0} (x) =\alpha_0 x +\alpha_1 (1-x)^2 +\frac{1}{2}\alpha_2 x(2-x)^2,
$$
so that the nonzero derivatives are
$$
\frac{\pa}{\pa x} e^x G^{\vert 1\rangle}_{\rho,0} (x)\bigg\vert_{x=0} = \alpha_0 -2\alpha_1 +2\alpha_2,\quad
\frac{\pa^2}{\pa x^2} e^x G^{\vert 1\rangle}_{\rho,0} (x)\bigg\vert_{x=0}  = 2\alpha_1 -4 \alpha_2,\quad
\frac{\pa^3}{\pa x^3} e^x G^{\vert 1\rangle}_{\rho,0} (x)\bigg\vert_{x=0} = 3\alpha_2.
$$
The reconstruction formula can then be written explicitly as
\begin{equation}\label{rekonstruktioesim}
\rho_{nn} = (\alpha_0 -2\alpha_1 +2\alpha_2 ) B^{1,0}_{n0} +(2\alpha_1 -4 \alpha_2) B^{1,0}_{n1} +3\alpha_2 B^{1,0}_{n2}.
\end{equation}

To obtain the matrix elements $B^{1,0}_{np}$, we first calculate the matrix $(A^{1,0}_{pn})_{p,n\in\N}$. It follows from elementary calculations that
$$
A^{1,0}_{pp} =p+1,\qquad A^{1,0}_{p,p+1} = -(2p+2),\qquad A^{1,0}_{p,p+2} =p+2,
$$
and $A^{1,0}_{pn}=0$ otherwise. In matrix form, this reads
$$
(A^{1,0}_{pn}) =
\left(\begin{array}{ccccccc}
1 & -2 &  2 &  0 &  0 &   0 & \cdots \\
0 &  2 & -4 &  3 &  0 &   0 & \cdots \\
0 &  0 &  3 & -6 &  4 &   0 & \cdots \\
0 &  0 &  0 &  4 & -8 &   5 & \cdots \\
0 &  0 &  0 &  0 &  5 & -10 & \cdots \\
0 &  0 &  0 &  0 &  0 &   6 & \cdots \\
\vdots & \vdots  & \vdots & \vdots & \vdots & \vdots &\ddots  \\
\end{array}\right)
$$
\end{example}
The inverse matrix can now be calculated, and we obtain
$$
(B^{1,0}_{np}) =
\left(\begin{array}{ccccccc}
1 &           1 & \frac{2}{3} & \frac{1}{4} & -\frac{2}{15} & -\frac{31}{72} & \cdots  \\ 
0 & \frac{1}{2} & \frac{2}{3} & \frac{5}{8} &  \frac{7}{15} & \frac{37}{144} & \cdots \\ 
0 &           0 & \frac{1}{3} & \frac{1}{2} &  \frac{8}{15} &  \frac{17}{36} & \cdots \\
0 &           0 &           0 & \frac{1}{4} &   \frac{2}{5} &  \frac{11}{24} & \cdots \\
0 &           0 &           0 &           0 &   \frac{1}{5} &    \frac{1}{3} & \cdots \\
0 &           0 &           0 &           0 &             0 &    \frac{1}{6} & \cdots \\
\vdots & \vdots & \vdots & \vdots & \vdots & \vdots & \ddots 
\end{array}\right)
$$
Inserting the appropriate elements into equation \eqref{rekonstruktioesim}, we find that 
\begin{eqnarray*}
\rho_{00} &=&  1\cdot(\alpha_0 -2\alpha_1 +2\alpha_2 ) + 1\cdot(2\alpha_1 -4\alpha_2) +\frac{2}{3}\cdot 3\alpha_2 =\alpha_0,\\
\rho_{11} &=& 0\cdot(\alpha_0 -2\alpha_1 +2\alpha_2 ) +\frac{1}{2}\cdot (2\alpha_1 -4\alpha_2) +\frac{2}{3}\cdot 3\alpha_2 =\alpha_1,\\
\rho_{22} &=& 0\cdot(\alpha_0 -2\alpha_1 +2\alpha_2 ) +0\cdot (2\alpha_1 -4\alpha_2) +\frac{1}{3}\cdot 3\alpha_2=\alpha_2,
\end{eqnarray*}
so that the formula does indeed give the correct values for the matrix elements.

\section{Discussion}\label{discussion}
In this article we have considered the problem of reconstruction the unknown state of a quantum system from the measurement statistics of phase space observables generated by the number states. The two reconstruction formulas we have derived are dealing with two different scenarios and, as such, have obviously different advantages and disadvantages.

In the first case, the reconstruction requires measurements of all of the observables $\mathsf{E}^{\vert s\rangle}$. From the practical point of view, this is of course impossible for many reasons. First of all, in the eight-port homodyne detection scheme, for the measurement of $\mathsf{E}^{\vert s\rangle}$, a parameter field needs to be prepared in the number state $\vert s\rangle\langle s\vert$. At the present, this is possible only for small values of $s$. Nevertheless, it might be reasonable to expect that future progress could allow sufficiently large number state preparations, so that the reconstruction would be possible with adequate precision. An advantage of this method is that one only needs to measure the observables near the origin, and the whole phase space does not need to be scanned.

The second method uses a single obervable $\mathsf{E}^{\vert s\rangle}$ and thus the problem of number state generation for arbitraily large values of $s$ is removed. This time the region of phase space needed is considerably larger, since, in principle, the method involves integrals over the whole space. This is of course an idealization which can not be considered in practice. From the numerical point of view, this method seems to be very manageable. Suppose that one measures the observable $\mathsf{E}^{\vert s\rangle}$ for some $s$. Using polar coordinates for the resulting phase space ditribution, one can then integrate the distribution with respect to the angle variable over $[0,2\pi)$ to obtain the quantity $G^{\vert s\rangle}_{\rho,0}$. If the density matrix is assumed to be finite, say, an $N\times N$-matrix, then $e^{r^2}G^{\vert s\rangle}_{\rho,0}(r)$ is a polynomial of order $2N+2s-2$. Choosing a sufficiently large $N\in\N$, one can thus fit such a polynomial into $e^{r^2}G^{\vert s\rangle}_{\rho,0}(r)$. This also fixes the size of the approximative density matrix. In view of the reconstruction formula, this assumption of finiteness has the crucial consequence that only the inverses of $N$ \emph{finite} matrices are needed. Similarly, the quantities $G^{\vert s\rangle}_{\rho,l}$ are needed only for $l=0,\ldots,N-1$.

\

\noindent {\bf Acknowledgment.} We wish to thank Pekka Lahti for useful comments on the manuscript. J. K. was supported by Emil Aaltonen Foundation and Finnish Cultural Foundation during the preparation of the manuscript. J. S. was supported by Turku University Foundation.

\end{document}